\documentclass{article}

\usepackage{arxiv}

\usepackage[utf8]{inputenc} 
\usepackage[T1]{fontenc}    
\usepackage{hyperref}       
\usepackage{url}            
\usepackage{booktabs}       
\usepackage{nicefrac}       
\usepackage{microtype}      
\usepackage{lipsum}
\usepackage{fancyhdr}       
\usepackage{doi}

\usepackage{amsmath}
\usepackage{amsfonts}
\usepackage{amssymb}
\usepackage{amsthm} 

\usepackage{graphics}

\usepackage{cancel}

\usepackage{tikz} 
\usepackage{pgfplots}

\usetikzlibrary{scopes}
\usetikzlibrary{arrows.meta}

\usepgfplotslibrary{polar}
\usepgfplotslibrary{colormaps}

\usetikzlibrary{external}

\newcommand{\kk}{k}

\newcommand{\cjlk}{{c}_{j,l,\kk}}
\newcommand{\phijlk}{{\varphi}_{j,l,\kk}}
\newcommand{\phihatjlk}{{\hat{\varphi}}_{j,l,\kk}}
\newcommand{\phijlkc}{{\varphi}^{*}_{j,l,\kk}}

\title{Curvelet-based Model for the Generation of Anisotropic Fractional
Brownian Fields}

\author{ \href{https://orcid.org/0000-0002-4411-3635}{Marcos V. C. Henriques} \\
	Departamento de Ci\^encias Exatas e Tecnologia da Informa\c{c}\~ao\\
	Universidade Federal Rural do Semi-\'Arido\\
	Angicos, Brazil \\
	\texttt{viniciuscandido@ufersa.edu.br} \\
}

\hypersetup{
	pdftitle={Curvelet-based Model for the Generation of Anisotropic Fractional
Brownian Fields},
	pdfauthor={Marcos Vinícius Cândido Henriques},
	pdfkeywords={curvelets, fractional brownian motion, fractal models},
}

\begin{document}

\maketitle

\begin{abstract}
We propose a curvelet-based model for the generation of Anisotropic
Fractional Brownian Fields, that are suited to model systems with
orientation-dependent self-similar properties. The synthesis procedure
consists of generating coefficients in the curvelet space with zero-mean
Gaussian distribution. This approach allows the representation of
natural systems having stochastic behavior in some degree and also
obeying to a given angular distribution of correlations. Examples
of such systems are found in heterogeneous geological structures,
in anisotropic materials and in complex disordered media.
\end{abstract}


\section{Introduction}

Systems with strong anisotropic properties are very frequent in nature,
but unfortunately they present serious difficulties in their characterization
and mathematical description. This is the case of complex systems
in geology, transport phenomena in material science, wave propagation
in disordered media and radiative transfer in systems that do not
have a spherical symmetry.

An example with technological application is related to the problem
of petroleum exploration. In this problem, the main used technique
is related to the phenomenon of scattering of seismic waves in geological
structures with anisotropic inhomogeneities in all scales. To take
into account the fluctuations of the physical quantities of these
structures it is necessary to describe their properties in a statistical
way. In these systems, the geological data are best modeled by fractal
geometries as the ones created by Mandelbrot \cite{mandelbrot1968}.
For example, it has been reported by several authors that well log
data from petroleum reservoirs show long-range correlation behavior,
which is a characteristic feature of fractal processes like the Fractional
Brownian Motion \cite{hewett1986,sahimi2005,sahimi2011,hardy1994}.
It has been also reported \cite{hansen2011} that long range correlations
in unconsolidated sandstones and porous media can be understood as
the consequence of extreme dynamics restructuring processes occurred
during the geological evolution. The onset of anisotropy in these
geological systems can be related to the influence of the gravitational
field and to macroscopic material flows. Characterizing long-range
correlations in these systems, such as porous media described by the
porosity logs, may prove useful for an accurate interpretation of
geological data.

When analysing layered rocks and geological fields formation due to
stratification process, we can verify that geometrical and/or transport
properties may be characterized in terms of their anisotropy degree
\cite{sahimi_94,Dullien_74,peterking1996}. Indeed, the probability
of occurrence of a given porosity (or permeability) gradient is strongly
dependent on the orientation of the rock properties. Consequently,
these systems are characterized by highly anisotropic correlations.
Studies in other areas deal with similar phenomena. For example, Ponson
et al \cite{ponson2006} have studied the self-affine properties of
anisotropic fracture surfaces. Jennane et al \cite{Jen_Maj_Lem_01}
have observed anisotropic correlation in images of bone X-ray tomographic
microscopy projections.

Since the early work of Mandelbrot \cite{mandelbrot1968}, the Fractional
Brownian Motion model (FBM) is largely being used to treat a wide
variety of natural phenomena having self-similarity and long-range
correlations. In particular, the fractal analysis study applied to
bidimensional fields has proved to be very useful to describe some
physical properties such as roughness and porosity. However, the standard
FBM model is only suitable for describing materials and media that
show isotropic symmetry. In that context, generalizations of the FBM
to anisotropic models have been proposed in last years. Kamont \cite{kamont1996}
studied the Fractional Anisotropic Wiener Field. Bonami and Estrade
\cite{bonami2003} analysed the properties of various models of anisotropic
Gaussian fields and proposed a new procedure to characterize the
anisotropy of such fields. Tavares and Lucena \cite{tavares2003}
and Heneghan \cite{heneghan1996} have implemented wavelet-based models
for anisotropic hypersurfaces. Bierm\'e and Richard \cite{bierme2006}
have applyied Radon Transform to estimate the anisotropy of Fractional
Brownian Fields. Xiao \cite{xiao2009} has studied sample path properties
of such anisotropic fields. It will be desirable to generalize these
concepts and ideas in simple and automatic ways that can allow sparse
representations of highly complex anisotropic fields.

In this paper, we introduce a new method for generating 2-D Anisotropic
Fractional Brownian Fields (AFBF) based on the Curvelet Transform.
This is a new multiscale transform with strong directional character
that provides an optimal representation of objects that have discontinuities
along edges \cite{candes2000,candes2002,candes2004}. The curvelets
are localized not only in the spatial domain (location) and frequency
domain (scale), but also in angular orientation, which is a step ahead
compared to Wavelet Transform \cite{mallat}. This directional feature
is the one we use to achieve the anisotropy.

\section{The Fractional Brownian Motion}

We start with the model of Fractional Brownian Motion proposed by
Mandelbrot \cite{mandelbrot1968}. An isotropic two-dimensional Fractional
Brownian field (FBF) $B_{H}(u)$, $u\in\mathbb{R}^{2}$, with Hurst
index $H$ taking values in $(0,1)$, is defined by the correlation
function
\begin{equation}
\mbox{E}\left[B_{H}(u)B_{H}(v)\right]=C_{H}\left(\left|u\right|^{2H}+\left|v\right|^{2H}+\left|u-v\right|^{2H}\right),\forall u,v\in\mathbb{R}^{2},\label{eq:defFBM2D}
\end{equation}
where $\textrm{E}\left[\cdot\right]$ is the expectation operator
and $C_{H}$ is a constant depending on $H$. $B_{H}(u)$ is not a
stationary process, but its increments form a stationary, zero-mean
gaussian process, with variance depending only on the distance $\Delta u$:
\[
\mbox{E}\left[\left|B_{H}(u+\Delta u)-B_{H}(u)\right|^{2}\right]\varpropto\Delta u^{H}.
\]

It follows from \eqref{eq:defFBM2D} that $B_{H}$ is a self-similar
field:

\[
B_{H}(\lambda u)\overset{d}{=}\lambda^{H}B_{H}(u),
\]
where $\overset{d}{=}$ means equal in distribution, and $\lambda>0$
is a constant. It can be proved that $B_{H}$ has an average spectral
density of the form:
\begin{equation}
S\left(\xi\right)\varpropto\left|\xi\right|^{-2H-2}\label{eq:FBMSpectralDensity}
\end{equation}
in which $\xi=(\xi_{1},\xi_{2})$ are the frequency coordinates in
the Fourier domain. As proposed in \cite{bonami2003}, anisotropic
fields with stationary increments can be defined by taking orientation-dependent
spectral densities of the form:

\[
S\left(\xi\right)\varpropto\left|\xi\right|^{-2H_{\theta}-2}
\]
where the Hurst index $H_{\theta}$ now depends on the direction of
$\xi$, labeled by the index $\theta$. We consider $H$ a parameter
related to the roughness and regularity of the surface. So, in the
case of anisotropy, we would expect that $H_{\theta}$ would be related
to the self-similarity properties of the field on a given direction.
A 1-D process obtained by selecting any straight line of an isotropic
FBF $B_{H}$ is a FBM with Hurst index $H$. However, contrary to
what one would expect, the estimation of the orientation-dependent
Hurst index $H$ in an anisotropic Gaussian field cannot be performed
simply by analysing sample lines of the field, as in the isotropic
case, since the regularity along a line do not appear to be dependent
on the direction. In order to measure the anisotropy dependence of
Hurst exponent $H_{\theta}$, Bonami and Strade \cite{bonami2003}
developed a procedure, the Directional Average Method (DAM), which
consists in computing the average over all the lines orthogonal to
$\theta$. The resulting 1-D process, a Fractional Brownian motion
(FBM), will have a Hurst index equal to $H_{\theta}+1/2$ (in fact,
they have proved this to be true for the critical H\"{o}lder exponent,
which in turn corresponds to the Hurst index in FBM paths \cite{bierme2006}).
We will use a modified version of this procedure later to test accuracy
of our curvelet-based method of synthesis.

\section{The Curvelet Transform}

The Curvelet transform is a recent multiscale analysis developed by
Cand\`{e}s and Donoho \cite{candes2000}. A curvelet at scale $2^{-j}$
is an oriented object whose support is the rectangle of width $2^{-j}$
and length $2^{-j/2}$ that obeys the parabolic scaling relation \textit{width}
$\approx$ \textit{length}$^{2}$ \cite{candes2006}. The curvelets
are described by a triple index $j$ (scale), $l$ (orientation) and
$k$ (spatial location). The basic curvelets elements are obtained
by rotations and translations of a specific basis function $\phi_{j}\in L^{2}(\mathbb{R}^{2})$,
called curvelet mother, that depends only on the scale index $j$
and is defined in the Fourier domain by the window function \cite{candes2000}:

\begin{equation}
\hat{\phi}_{j}(r,\omega)=2^{-3j/4}W\left(2^{-j}r\right)V\left(\frac{2^{\left\lfloor j/2\right\rfloor }\omega}{2\pi}\right),\label{equ1}
\end{equation}
 where we use the polar coordinates $(r,\omega)$ in the Fourier domain,
and $\left\lfloor j/2\right\rfloor $ is the integer part of $j/2$.
$W$ and $V$ are the ``radial window'' and the ``angular window'',
respectively. These two functions are smooth, non-negative and real-value,
with $W$ taking positive real arguments. $W$ and $V$ restrict the
support $\hat{\phi_{j}}$ to a polar wedge that is symmetric respect
to zero.

Let us define the window $u_{j,l}=\hat{\phi}_{j}\left(R_{\theta_{j,l}}\xi\right)$,
where $\xi\in\mathbb{R}^{2}$ is the cartesian coordinates in the
Fourier domain and $R_{\theta_{j,l}}$ is the rotation matrix by $\theta_{j,l}$
radians. The family of curvelets $\phi_{j,l,k}$ is defined in the
Fourier domain as:
\[
\hat{\phi}_{j,l,k}=u_{j,l}e^{i\left\langle b_{k}^{j,l},\xi\right\rangle },
\]
at scale $2^{-j}$, orientation $\theta_{j,l}$, and position $b_{k}^{j,l}=R_{\theta_{j,l}}^{-1}(2^{-j}k_{1},2^{-j/2}k_{2})$,
where $k=(k_{1},k_{2})\in\mathbb{Z}^{2}$. The curvelet transform
of a function $f\in L^{2}(\mathbb{R}^{2})$ is given by the convolution
integral: 
\begin{equation}
\cjlk=\left\langle f,\phijlk\right\rangle =\int_{\mathbb{R}^{2}}f(x)\phijlkc(x)dx.\label{eq:transform_coefficients}
\end{equation}
where $\overline{\phi}$ denotes the complex conjugate of $\phi$.
In this equation the coefficients $c_{j,l,k}$ are interpreted as
the decomposition of $f$ into a basis of curvelets functions $\phijlk$
\cite{candes2005a}. 

\begin{figure}
\begin{centering}
\scalebox{.75}{\includegraphics{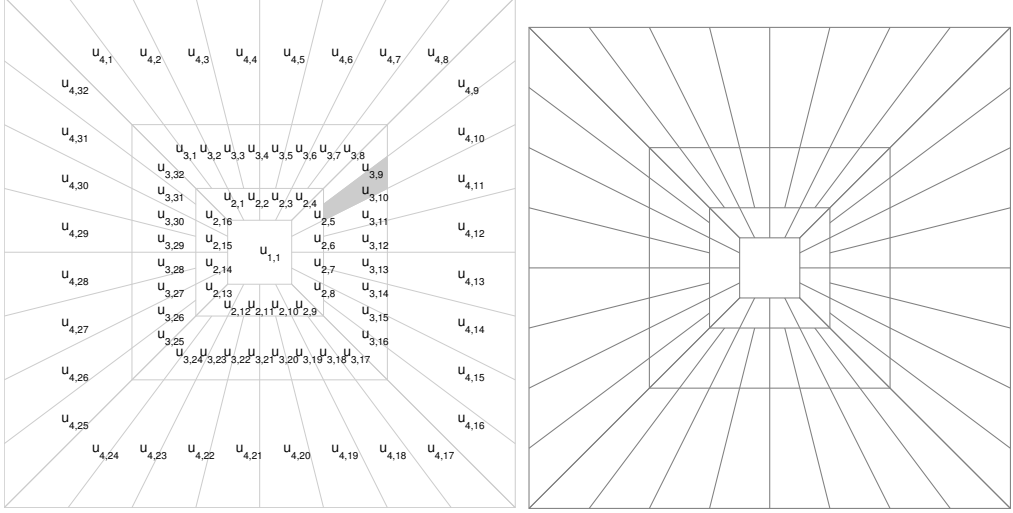}}
\begin{tikzpicture}[y=.2cm, x=.2cm]
    	\foreach \x in {-12,-4,...,12}
     		\draw[gray] (0,0) -- (\x,-16);
		\foreach \x in {-12,-4,...,12}
     		\draw[gray] (0,0) -- (\x,16);
		\foreach \y in {-12,-4,...,12}
     		\draw[gray] (0,0) -- (-16,\y);
		\foreach \y in {-12,-4,...,12}
     		\draw[gray] (0,0) -- (16,\y);

		\filldraw[fill=white, draw=gray] (-4,-4) rectangle (4,4); 

		\foreach \x in {-16,-8,...,16}
     		\draw[gray] (0,0) -- (\x,-16);
		\foreach \x in {-16,-8,...,16}
     		\draw[gray] (0,0) -- (\x,16);
		\foreach \y in {-16,-8,...,16}
     		\draw[gray] (0,0) -- (-16,\y);
		\foreach \y in {-16,-8,...,16}
     		\draw[gray] (0,0) -- (16,\y);

		\filldraw[fill=white, draw=gray] (-2,-2) rectangle (2,2); 

		\draw[gray] (-16,-16) rectangle (16,16); 
		\draw[gray] (-8,-8) rectangle (8,8); 


\end{tikzpicture}
\caption{A basic discrete curvelet tiling with the first 4 scales in the Fourier
domain. Each $u_{j,l}$ represents a window function at scale $2^{-j}$
and orientation $\theta_{j,l}$. A sample support for a given window
$u_{j,l}$ is marked with gray.\label{fig:Tilling}}
\end{centering}
\end{figure}

The discrete version of this transform is done by choosing a discrete
tiling in the Fourier domain with pseudo-polar supports for the window
functions $u_{j,l}$, which is more adapted to cartesian arrays \cite{candes2006}.
Figure \ref{fig:Tilling} shows an example of such a discrete curvelet
tiling, with each wedge being linked to its corresponding window function
$u_{j,l}$. The discrete curvelet transform consists, roughly speaking,
in obtaining inner products in the Fourier domain, since, based on
Plancherel's Theorem, we have $\cjlk=\left\langle f,\phijlk\right\rangle =\left\langle \hat{f},\phihatjlk\right\rangle $.
Therefore we can recover the original signal by using the reconstruction
formula
\begin{equation}
f=\sum_{j,l,k}\cjlk\,\phijlk\label{eq:reconstruction_formula}
\end{equation}

\section{The Curvelet-based Model for Anisotropic Fractional Brownian Fields}
Wavelet-based methods for synthesis of two-dimensional FBF have been
proposed by Tavares \cite{tavares2003} and Heneghan et al \cite{heneghan1996}.
They show that the wavelet detail coefficientes $d_{j,k}$ at scale
$2^{-j}$ of a 2-D FBF with Hurst index $H$ have a zero-mean Gaussian
distribution with variance that varies as a power-law in scale with
the form:
\begin{equation}
\textrm{E}\left[\left|d_{j,k}\right|^{2}\right]=C_{H}^{\psi}2^{-j(2H+2)},\label{eq:FBF_wavelet_coefs_variance}
\end{equation}

where $C_{H}^{\psi}$ is a constant depending on $H$ and the wavelet
$\psi$ used, and $k\in\mathbb{Z}^{2}$ is a index related to spacial
location. So, in order to synthesize a 2-D FBF, one has only to generate
a series of coefficients with Gaussian distribution and the given
variance at each scale, and then transform to the real space.

\begin{figure}
\noindent \begin{centering}
\includegraphics[scale=0.35]{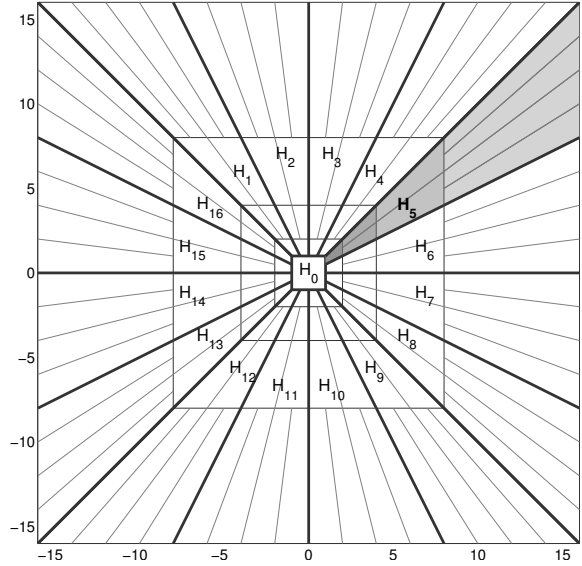}
\par\end{centering}
\caption{A sample set of wedges in the Fourier plane associated with a given
$H_{m}$ index is marked with gray.\label{fig:tilling-H}}
\end{figure}

We construct the anisotropic FBF with the curvelet transform in the
following way. We introduce an angular variable by dividing the $2\pi$
angle of the plane into $N$ identical angular sectors which are identified
by a variable $m$. For each angle index $l$ and scale $2^{-j}$,
$j>1$, we generate a matrix of coefficients $c_{j,l,k}$ with zero-mean
Gaussian distribution and variance 
\begin{equation}
\textrm{E}\left[\left|c_{j,l,k}\right|^{2}\right]=C_{H_{m}}^{\phi}2^{-j(2H_{m}+2)},\label{eq:AFBF_curvelet_coefs_variance}
\end{equation}

where now the Hurst index, $H_{m}$, depends on the index $m=1,2,...,N$,
related to the orientation, which in turn is independent of scale
and can assume one of the $N$ integer values. For each angular sector
there is only one value for $m$. $C_{H_{m}}^{\phi}$ is a constant
depending on $H_{m}$ and on the curvelet $\phi$. For practical reasons,
we consider $C_{H_{m}}^{\phi}$ = constant, independently of $m$.
The relation among $m$, the curvelet indexes $j$ (scale) and $l$
(scale-dependent orientation) is 
\[
m=\left\lceil Nl/n_{j}\right\rceil 
\]

where $\left\lceil x\right\rceil $ denotes the smallest integer being
greater than or equal to $x$, and $n_{j}=8\cdot2^{\left\lceil j/2\right\rceil }$,
for $j>1$, is the number of available angles at the scale $2^{-j}$,
which is determined by the parameters of the curvelet transform. We
have set the first scale $j_{0}\equiv1$ and, by definition, $n_{1}\equiv1$,
that is to say, there is no distinct directions at the coarsest scale.
We define $N$ to be the number of available angles at second scale
($N\equiv n_{2}$), since this is the maximum possible number of angular
sectors we can split the discrete tiling of frequency space (see figure
\ref{fig:tilling-H}). In our case we use $N=n_{2}=16$, so $H_{m}$
can assume $16$ different values. For heuristic reasons, we have
set $H_{0}=\textrm{E}\left[H_{m}\right]$ to be the index associated
with the first scale. In figure \ref{fig:tilling-H} is shown a sample
set of wedges in the Fourier plane associated with a given $H_{m}$
index.

After generating the coefficients, all we have to do is to perform
the inverse curvelet transform as expressed by \eqref{eq:reconstruction_formula}
to get the anisotropic field.

\section{Results}

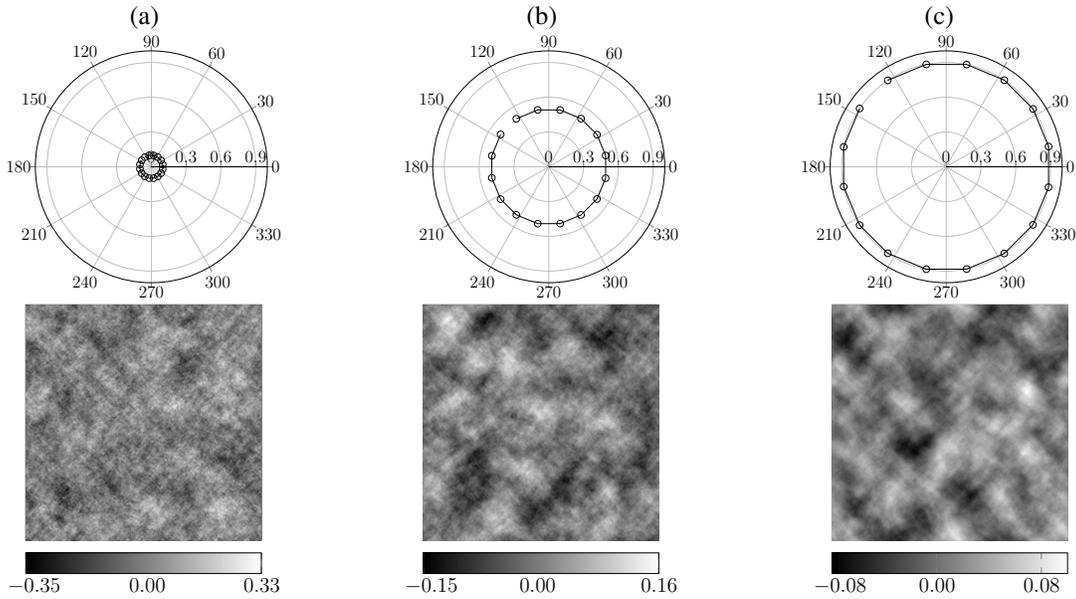
\begin{figure}[h]
\noindent \begin{centering}
\begin{minipage}[c][1\totalheight][t]{0.32\columnwidth}%
\begin{center}
(a)\\
\pgfplotsset{
every x tick 
label/.append 
style={font=\Large, yshift=0.0ex}} 

\pgfplotsset{
every y tick 
label/.append 
style={font=\Large, yshift=0.0ex}} 

\begin{tikzpicture}[scale=0.45]
\begin{polaraxis}[
   ytick={0,0.3,...,1},
   ymin=0, ymax=1,
]
\addplot [thick, black, mark=o, mark size=2.8] table [col sep=comma, x=theta, y=H] {Hdist_0_11_4e-170_1.dat}; 
\end{polaraxis}
\end{tikzpicture}\\
\begin{tikzpicture}[scale=0.55]
\begin{axis}[
		ticks=none,
		enlargelimits=false, 
		axis equal image,
		colormap/gray,
		colorbar horizontal,
	    point meta min=-0.35,
		point meta max= 0.33,
		colorbar style={
			anchor=north west,
        	xtick={-0.35,0,0.33},
        	xticklabel style={
            text width=2.5em,
            align=right,
            /pgf/number format/.cd,
                fixed,
                fixed zerofill
        }
    }
]
 \addplot graphics[xmin=0,xmax=1,ymin=0,ymax=1,zmin=0,zmax=.5]
   {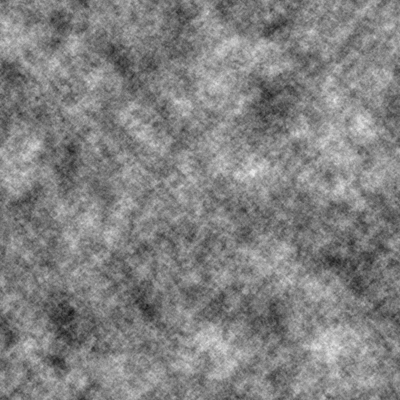};
\end{axis}
\end{tikzpicture}
\par\end{center}%
\end{minipage}%
\begin{minipage}[c][1\totalheight][t]{0.32\columnwidth}%
\begin{center}
(b)\\
\pgfplotsset{
every x tick 
label/.append 
style={font=\Large, yshift=0.0ex}} 

\pgfplotsset{
every y tick 
label/.append 
style={font=\Large, yshift=0.0ex}} 

\begin{tikzpicture}[scale=0.45]
\begin{polaraxis}[
   ytick={0,0.3,...,1},
   ymin=0, ymax=1,
]
\addplot [thick, black, mark=o, mark size=2.8] table [col sep=comma, x=theta, y=H] {Hdist_0_500_5.dat}; 
\end{polaraxis}
\end{tikzpicture}\\
\begin{tikzpicture}[scale=0.55]
\begin{axis}[
		ticks=none,
		enlargelimits=false, 
		axis equal image,
		colormap/gray,
		colorbar horizontal,
	    point meta min=-0.15,
		point meta max= 0.16,
		colorbar style={
			anchor=north west,
        	xtick={-0.15,0,0.16},
        	xticklabel style={
            text width=2.5em,
            align=right,
            /pgf/number format/.cd,
                fixed,
                fixed zerofill
        }
    }
]
 \addplot graphics[xmin=0,xmax=1,ymin=0,ymax=1,zmin=0,zmax=.5]
   {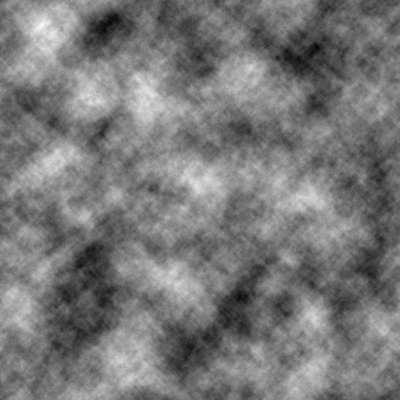};
\end{axis}
\end{tikzpicture}
\par\end{center}%
\end{minipage}%
\begin{minipage}[c][1\totalheight][t]{0.32\columnwidth}%
\begin{center}
(c)\\
\pgfplotsset{
every x tick 
label/.append 
style={font=\Large, yshift=0.0ex}} 

\pgfplotsset{
every y tick 
label/.append 
style={font=\Large, yshift=0.0ex}} 

\begin{tikzpicture}[scale=0.45]
\begin{polaraxis}[
   ytick={0,0.3,...,1},
   ymin=0, ymax=1,
]
\addplot [thick, black, mark=o, mark size=2.8] table [col sep=comma, x=theta, y=H] {Hdist_0_92_3e-160_9.dat}; 
\end{polaraxis}
\end{tikzpicture}\\
\begin{tikzpicture}[scale=0.55]
\begin{axis}[
		ticks=none,
		enlargelimits=false, 
		axis equal image,
		colormap/gray,
		colorbar horizontal,
	    point meta min=-0.08,
		point meta max= 0.1,
		colorbar style={
			anchor=north west,
        	xtick={-0.08,0,0.08},
        	xticklabel style={
            text width=2.5em,
            align=right,
            /pgf/number format/.cd,
                fixed,
                fixed zerofill
        }
    }
]
 \addplot graphics[xmin=0,xmax=1,ymin=0,ymax=1,zmin=0,zmax=.5]
   {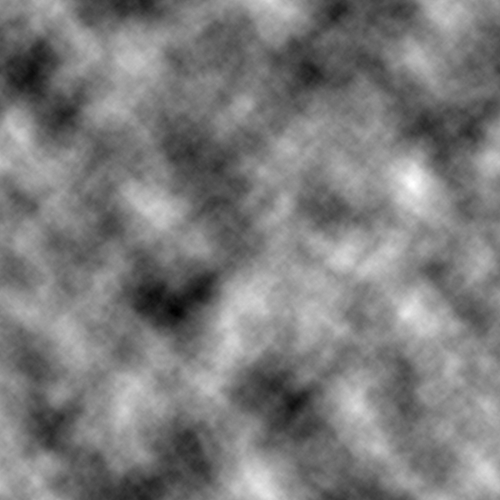};
\end{axis}
\end{tikzpicture}
\par\end{center}%
\end{minipage}
\par\end{centering}
\caption{Three examples of isotropic Fractional Brownian Fields obtained by
using the curvelet method. Top: the angular distribution of $H_{m}$
index used in the synthesis of each field. Bottom: the corresponding
synthesized fields. \label{fig:Synthesized-2-D-Gaussian-iso}}
\end{figure}

\begin{figure}[h]
\noindent \begin{centering}
\includegraphics{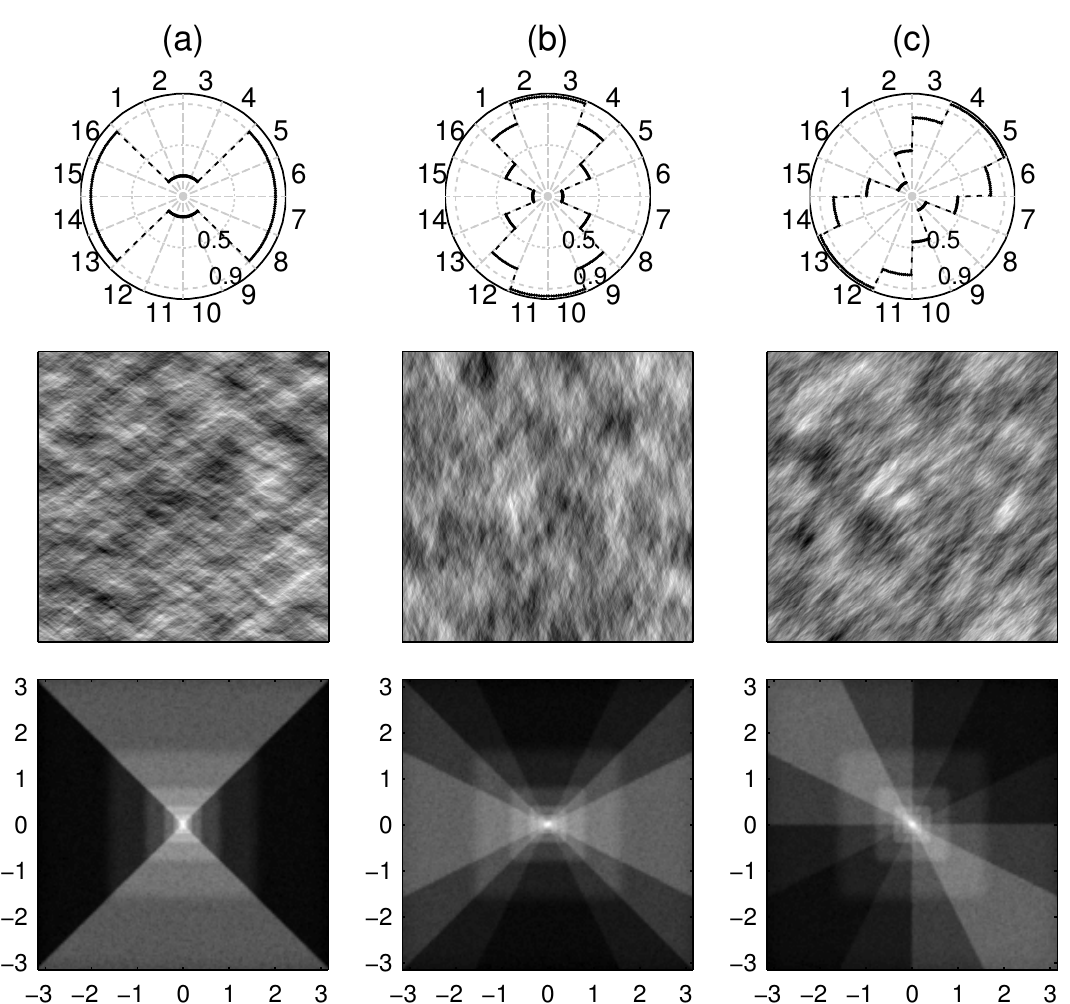}
\par\end{centering}
\caption{Three examples of anisotropic Fractional Brownian Fields obtained
by using the curvelet method. Top: the angular distribution of $H_{m}$
index used in the synthesis of each field. Middle: the corresponding
synthesized fields. Bottom: The corresponding spectral density.\label{fig:Synthesized-2-D-Gaussian}}
\end{figure}

Figure \ref{fig:Synthesized-2-D-Gaussian} shows three synthesized
isotropic fractional brownian fields. (a) presents an isotropic field
with Hurst index $0.1$, (b) presents an isotropic field with Hurst
index $0.5$ and (c) presents an isotropic field with Hurst index
$0.9$. Figure \ref{fig:Synthesized-2-D-Gaussian} shows three simulations
of anisotropic fractional brownian fields fields. The angular distribution
of the $H_{m}$ indexes used in each simulation is shown on top. (a)
we see a case with abrupt change of $H_{m}$ in the angular distribution,
from $0.2$ around the vertical direction to $0.9$ around the horizontal
direction. (b) presents an anisotropic field, with Hurst indexes ranging
from $0.1$ along east direction to $0.9$ along north direction.
We can see in this example that the generated field seems to have
anticorrelation along one axis (east) and correlation along other
(north). (c) presents an anisotropic field, with Hurst indexes ranging
from $0.1$ along a direction almost northwest to $0.9$ along almost
northeast. At the bottom is shown the logarithm of the power spectrum
illustrating the directional character of the synthesized fields.

\begin{figure}[h]
\noindent \begin{centering}
\includegraphics[scale=0.6]{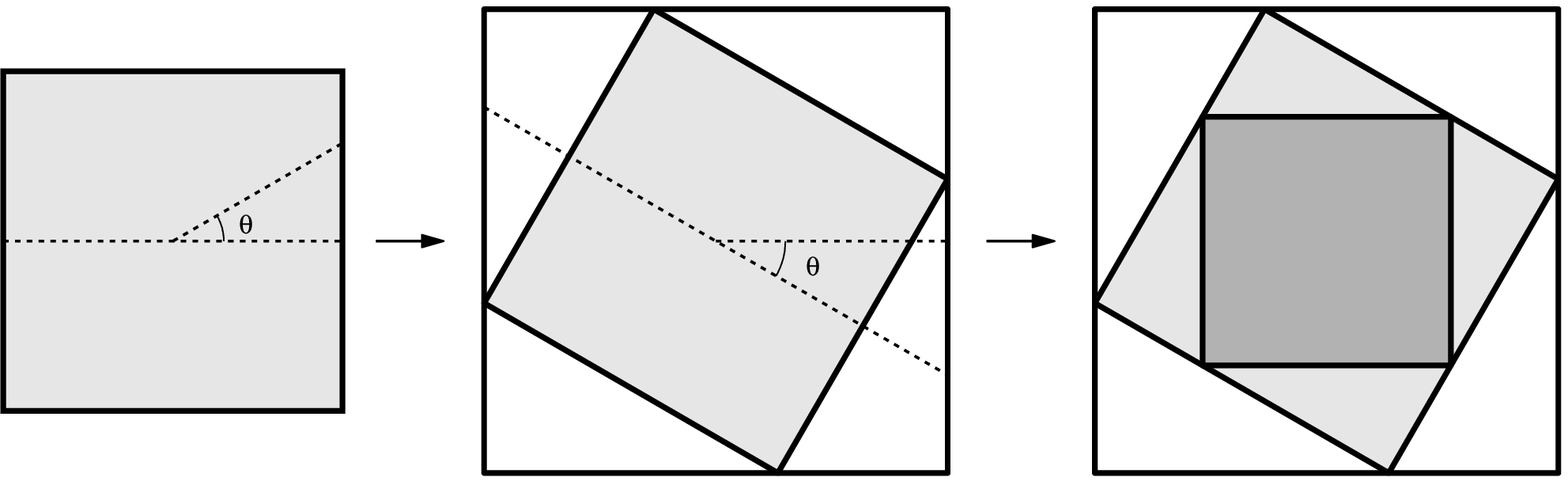}
\par\end{centering}
\caption{Scheme of rotation of the image that represents the field.\label{fig:image-rotation} }
\end{figure}

\begin{figure}[h]
\noindent \begin{centering}
\includegraphics{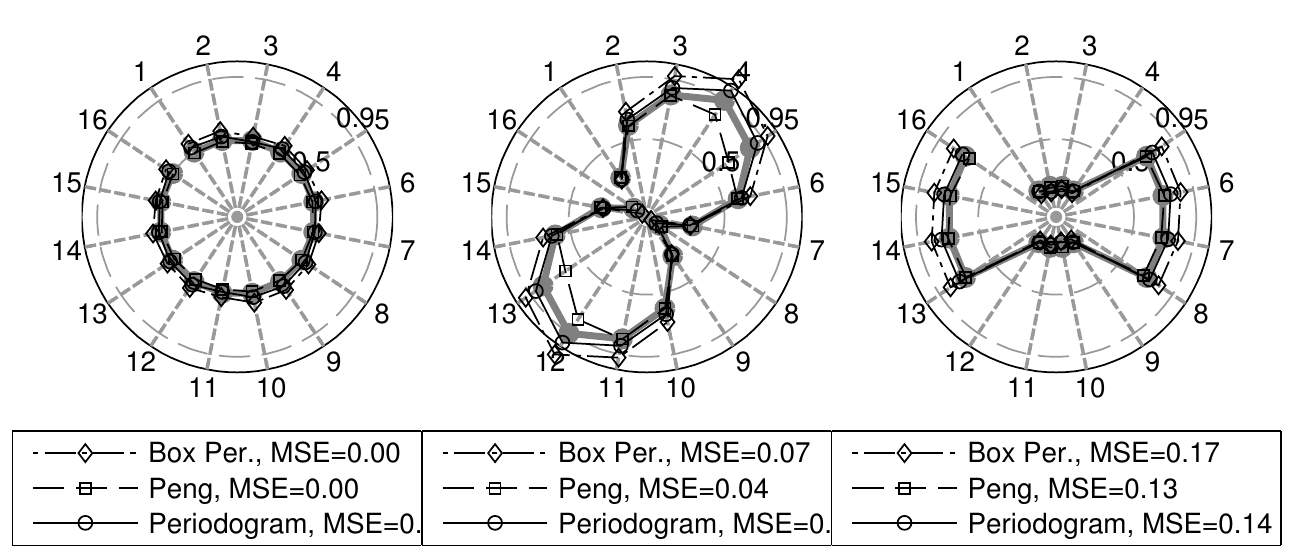}
\par\end{centering}
\caption{Anisotropy analysis by the Directional Average Method (DAM) of 50
samples with {\large{}$2000\times2000$} points. (MSE= Mean Squared
Error)\label{fig:Anisotropy-analysis} }
\end{figure}

For the evaluation of the quality of our synthesis procedure, we
must measure the Hurst index for each direction in the generated field
and to compare with the $H_{m}$ value used in the construction. For
this, we apply a modified version of the Directional Average Method
(DAM) proposed in \cite{bonami2003}. As said before, this method
was developed to estimate the orientation-dependent index $H$ in
anisotropic Gaussian fields. It consists in obtaining a 1-D signal
that is the average over all the lines orthogonal to a given direction
$\theta$. The estimated Hurst index of this signal must be about
$H_{\theta}+1/2$, where $H_{\theta}$ is the desired parameter to
characterize the field. In order to get the 1-D signal we proceed
as follows. The matrix representing the field is treated as a monochrome
image. We rotate this image by $\theta$ in a clockwise direction,
by using an image processing operator that includes a bicubic interpolation
\cite{keys_interpolation}, but others methods can also be applied.
The output, expressed as a matrix, is large enough to contain the
information of the entire rotated image, from which we extract an
inscribed submatrix whose columns represent an approximation to the
lines orthogonal to $\theta$ in the original data. This procedure
is ilustrated in figure \ref{fig:image-rotation}. So we can directly
get the output 1-D signal by taking the average of each column of
this submatrix. Finally, we estimate the Hurst index of the signal
by using a method of choice. This process is performed, in our case,
for each one of the $N$ angles used in the synthesis by curvelets.

Figure \ref{fig:Anisotropy-analysis} shows the mean of the estimated
Hurst parameters at $16$ directions of 50 generated samples with
$2000\times2000$ points. Three methods of FBM analysis were used.
The first one, proposed by Peng et al \cite{peng1994}, uses Detrended
Fluctuation Analysis (DFA) to estimate $H$. The others two, periodogram
method and boxed periodogram method, use the observation that the
spectral density of a FBM behaves like \eqref{eq:FBMSpectralDensity}.
For details of these last two methods we refer to \cite{taqqu1995}.
The FBM analysis method that gives smaller mean squared error is the
DFA-based method, as shown in the figure.

\section{Final Remarks}

We have proposed a curvelet-based scheme for generating Anisotropic
Fractional Brownian Fields with orientation-dependent self-similar
properties. It can create surfaces with an arbitrary angular distribution
of Hurst parameters, which is a step forward with respect to wavelet-based
approachs such as Tavares and Lucena method \cite{tavares2003}. Our
proposed algorithm gives good results even in case of abrupt change
of the orientation-dependent Hurst index $H_{m}$ in the angular distribution. 

We expect this method to be useful in the study of systems where long-range
dependent stochastic processes arise, as for example, in geostatistics
of large scale anisotropic porous media. All we have to do is to generate
curvelet coefficients with zero-mean Gaussian distributions and variances
obeying to \eqref{eq:AFBF_curvelet_coefs_variance}, and then perform
the inverse discrete curvelet transform what gives a field with the
angular distribution of $H$ exponents that was introduced in the
curvelet space. The inverse curvelet transform stage of our synthesis
method can be performed by fast discrete algorithms \cite{candes2006}.
The generated fractal surfaces have spectral densities following a
orientation-dependent power-law, in a way that resembles the basic
discrete curvelet tiling.

\section*{Acknowledgments}

The author wish to dedicate this work to the memory of prof. Liacir dos Santos Lucena, and gratefully acknowledge the support of the Universidade Federal do Rio Grande do Norte (UFRN) and Universidade Federal Rural do Semi-\'Arido (UFERSA).

\bibliographystyle{unsrtnat}

\bibliography{references}

\end{document}